\documentclass[twocolumn,amsmath,amssymb,prl]{revtex4}
\usepackage{color}
\usepackage[pdftex]{graphicx}
\usepackage{dsfont}

\newcommand{\dd}{\mathbf{d}}
\newcommand{\rr}{\mathbf{r}}

\newcommand{\rrho}{\boldsymbol{\rho}}
\newcommand{\kk}{\mathbf{k}}
\newcommand{\qq}{\mathbf{q}}
\newcommand{\pp}{\mathbf{p}}

\input epsf
\begin{document}
\title {Beliaev damping in quasi-2D dipolar condensates}
\author{Ryan M. Wilson$^{1}$ and Stefan Natu$^2$}
\affiliation{$^1$Department of Physics, The United States Naval Academy, Annapolis, MD 21402, USA}
\affiliation{$^2$Condensed Matter Theory Center and Joint Quantum Institute, Department of Physics, University of Maryland, College Park, MD 20742, USA}
\begin{abstract}
We study the effects of quasiparticle interactions in a quasi-two dimensional (quasi-2D), zero-temperature Bose-Einstein condensate of dipolar atoms, which can exhibit a roton-maxon feature in its quasiparticle spectrum.  Our focus is the Beliaev damping process, in which a quasiparticle collides with the condensate and resonantly decays into a  pair of quasiparticles.  Remarkably, the rate for this process exhibits a highly non-trivial dependence on the quasiparticle momentum and the dipolar interaction strength.  For weak interactions, the low energy phonons experience no damping, and the higher energy quasiparticles undergo anomalously weak damping.  In contrast, the Beliaev damping rates become anomalously large for stronger dipolar interactions, as rotons become energetically accessible as final states. Further, we find a qualitative anisotropy in the damping rates when the dipoles are tilted off the axis of symmetry.  Our study reveals the unconventional nature of Beliaev damping in dipolar condensates, and has important implications for ongoing studies of equilibrium and non-equilibrium dynamics in these systems. 
\end{abstract}
\maketitle

The quasiparticle picture of fluctuations and excited states in condensed matter systems is a fundamental modern paradigm.
Early investigations in this direction focused on superfluid $^4$He, which hosts very low energy quasiparticles at intermediate wave vectors, termed ``rotons''~\cite{Landau47,Feynman54,Feynman56}.  Rotons were first observed in neutron scattering experiments with $^4$He~\cite{Landau47,Cohen57,Henshaw61,Dietrich72}, and are now understood to emerge in strongly interacting superfluids due to strong, longer-range two-body correlations~\cite{Schneider71,Henkel10,Mottl12}.
  Bose-Einstein condensates (BECs) of atoms with large magnetic dipole moments, such as Cr, Er, or Dy, are unique in that they are predicted to support roton quasiparticles when confined to highly oblate, quasi-two dimensional (quasi-2D) geometries, despite remaining extremely dilute and weakly interacting compared to superfluid $^4$He~\cite{Santos03,Fischer06,Ronen07,Wilson08}.  Thus, mean-field theories typically provide  good descriptions of these systems~\cite{Lahaye09,Baranov12}, despite their treatment of quasiparticles as free, non-interacting excitations.  
Here, we systematically step beyond the mean-field approximation, and study the effect of quasiparticle interactions on the damping of collective excitations in quasi-2D dipolar condensates, finding non-trivial effects beyond the free quasiparticle picture.


In  1958, Beliaev first presented a theory of the Bose-condensed state that includes quasiparticle interactions, showing how they manifest as effective condensate-mediated processes~\cite{Beliaev58a,Beliaev58b}.  An important consequence of such interactions is the damping of quasiparticle motion, resulting in finite lifetimes for collective condensate excitations.  Beliaev specialized to the case of isotropic, short-range (contact) interactions, which is relevant for alkali atom condensates~\cite{Dalfovo99}.  A number of subsequent works have following along these lines~\cite{Liu97,Giorgini98,Giorgini00,Capogrosso10,Zheng14,Pixley15}, and there is notable agreement with experimental work~\cite{Jin97,Katz02}.  However, despite a growing interest in the experimental study of quantum many-body physics with dipolar atoms~\cite{Griesmaier05,Lahaye07,Lu11,Lu12,Bismut12,Aikawa12,Kadau15arXiv} and polar molecules~\cite{Ni08,Deiglmayr08,Carr09,Aikawa09,Takekoshi14}, a systematic theoretical understanding of beyond mean-field effects, such as quasiparticle damping, is lacking for these systems.

In this Letter, we present a theory describing these effects in a quasi-2D dipolar BEC, and find a number of striking results.   When the dipolar interactions are weak, the damping rates are anomalously small, being significantly less than those of a gas with purely contact interactions of equal strength. In contrast, when the dipolar interactions are stronger and rotons begin to emerge in the quasiparticle spectrum, the Beliaev damping rates acquire anomalously large values, though the rotons themselves remain undamped.  For all interaction strengths, the low energy phonon modes are immune to damping \cite{Natu13a}.  Additionally, the dipolar interactions can be made strongly anisotropic in the quasi-2D geometry~\cite{Ticknor11}.  In this case, the Beliaev damping rates acquire qualitatively different character depending on the direction of quasiparticle propagation; this feature has no analog in conventional superfluids.  Our results mark an important step towards understanding the physics of dipolar condensates beyond the mean-field approximation, and have important implications for both the equilibrium and non-equilibrium properties of these novel superfluids.

In the grand canonical ensemble, the  Bose gas Hamiltonian is
\begin{align}
\label{Ham}
\hat{H} &= \int d\rr \, \hat{\psi}^\dagger (\rr) \left( \frac{\pp^2}{2m} + U(\rr) - \mu \right) \hat{\psi}(\rr) \nonumber \nonumber \\
&+ \frac{1}{2} \int d\rr \int d\rr^\prime \hat{\psi}^\dagger(\rr) \hat{\psi}^\dagger(\rr^\prime) V(\rr-\rr^\prime) \hat{\psi}(\rr^\prime) \hat{\psi}(\rr).
\end{align}
Here, $m$ is the atomic mass, $U(\rr)$ is the external potential, $\mu$ is the chemical potential of the gas, and $\hat{\psi}(\rr)$ ($\hat{\psi}^\dagger(\rr)$) is the Bose annihilation (creation) operator.  For fully polarized dipoles with dipole moments $\mathbf{d}$, the two-body interaction potential is $V(\rr) = d^2 (1 - 3 \cos^2 \theta ) / | \rr |^3$, 
where $\theta$ is the angle between $\rr$ and $\mathbf{d}$.  

At ultracold temperatures, the dilute Bose gas can be described by a mean-field theory with a condensate order parameter $\phi(\rr) = \langle \hat{\psi}(\rr) \rangle$, which evolves under the equation of motion,
\begin{align}
\label{Heis}
i \hbar \dot\phi(\rr) &= \left( \frac{\pp^2}{2m} + U(\rr) - \mu \right) \phi(\rr) \nonumber \\
 &+ \int d\rr^\prime V(\rr-\rr^\prime) \langle \hat{\Psi}^\dagger(\rr^\prime) \hat{\Psi}(\rr^\prime) \hat{\Psi}(\rr)  \rangle.
\end{align}
Under the decomposition $\hat{\Psi}(\rr) = \phi(\rr) + \hat{\varphi}(\rr)$, where $\hat{\varphi}(\rr)$ annihilates non-condensed atoms, $ \langle \hat{\Psi}^\dagger(\rr^\prime) \hat{\Psi}(\rr^\prime) \hat{\Psi}(\rr)  \rangle \simeq n(\rr^\prime) \phi(\rr) + \tilde{n}(\rr^\prime,\rr) \phi(\rr^\prime)$, 
where $n(\rr) = |\phi(\rr)|^2 + \tilde{n}(\rr,\rr)$ is the total density of the gas and $\tilde{n}(\rr^\prime,\rr) = \langle \hat{\varphi}^\dagger (\rr^\prime)  \hat{\varphi}(\rr) \rangle$ is the non-condensate density matrix.  We work in the Popov approximation, and omit the anomalous density matrix $\tilde{m}(\rr^\prime,\rr) =  \langle \hat{\varphi}(\rr^\prime) \hat{\varphi}(\rr) \rangle$ from the theory~\cite{Popov72}. In the perturbative framework we employ, the Beliaev damping rates are insensitive to this approximation~\cite{Giorgini98,Giorgini00}.

Small amplitude condensate oscillations can be modeled as perturbations $\delta \phi(\rr)$ about the stationary state of Eq.~(\ref{Heis}), denoted $\phi_0(\rr)$.  We obtain equations of motion for these condensate oscillations by inserting $\phi(\rr) = \phi_0(\rr) + \delta \phi(\rr)$ into Eq.~(\ref{Heis}) and linearizing about $\delta$.  If the couplings between $\delta \phi(\rr)$ and the non-condensate density $\tilde{n}$ are ignored, this procedure reproduces the Bogoliubov free-quasiparticle description of small amplitude condensate oscillations.
This description, however, is inadequate to describe the \emph{damping} of condensate oscillations.  To correct this, we couple the condensate oscillations, which take the form of Bogoliubov quasiparticles, to the non-condensate atoms perturbatively in $\delta$,  following the procedures of Refs.~\cite{Giorgini98,Giorgini00,Natu13a,Natu13b}.
We obtain eigenfrequencies $\omega^\prime = \omega + \delta \omega$, where $\omega$ are the bare (non-interacting) quasiparticle frequencies and $\delta \omega$ are  frequency shifts that arise due to quasiparticle interactions.  The imaginary part of $\delta \omega$ corresponds to a damping rate for condensate oscillations.  At $T=0$, this is a Beliaev process, which involves the resonant decay of a quasiparticle into a pair of quasiparticles under the constraints of energy and momentum conservation~\cite{Beliaev58a,Beliaev58b}.  The relevant damping rate is thus $\gamma_{\mathrm{B}} = \mathrm{Im}[\delta \omega]_{T=0}$.  This perturbative scheme remains valid for large damping rates, as long as the non-condensate density remains small.  

We restrict our study to the quasi-2D regime, where the atoms are free to move in-plane but are tightly confined in the axial direction by an external potential $U(\rr) = m \omega_z^2 z^2 / 2$.  If $\hbar \omega_z$ is the dominant energy scale in the system, to a good approximation all atoms occupy the single-particle ground state in the $z$-direction, $\chi(z) = \exp [ -z^2 / 2 l_z^2 ] / \sqrt{\pi} l_z^{1/4}$, where $l_z = \sqrt{\hbar / m \omega_z }$.  An effective quasi-2D theory is obtained by separating all bosonic fields into this axial wave function and integrating the $z$-coordinate from the theory~\cite{Petrov00}.  Below, we rescale all lengths in units of $l_z$ and all energies in units of $\hbar \omega_z$.  
The condensate order parameter becomes $\phi(\rr) = \sqrt{n_0} \chi(z)$ where $n_0$ is the uniform areal condensate density, and the condensate oscillations take the form $\delta \phi (\rr) = \chi(z) \sum_\pp ( u_\pp e^{i (\pp \cdot \rrho - \omega_\pp^\prime t) } + v^*_\pp e^{-i (\pp \cdot \rrho - \omega_\pp^\prime t) })$, where $\rrho$ and $\pp$ are in-plane spatial and momentum coordinates, respectively. The coefficients $u_\pp$ and $v_\pp$ are the Bogoliubov quasiparticle amplitudes, given by $u_\pp = \sqrt{\varepsilon_\pp / 2 \omega_\pp + 1}$ and $v_\pp = - \mathrm{sgn} [\tilde{V}(\pp) ] \sqrt{\varepsilon_\pp / 2 \omega_\pp - 1}$, where $\varepsilon_\pp = p^2/2 + n_0 g_d \tilde{V}(\pp)$.  The bare quasiparticle spectrum is
\begin{align}
\omega_\pp = \sqrt{\frac{p^2}{2}\left( \frac{p^2}{2} + 2 g_d n_0 \tilde{V} \left( \frac{\pp}{\sqrt{2}} \right) \right)},
\end{align}
 where $g_d = \sqrt{8 \pi} d^2 / 3 $ is the quasi-2D dipolar interaction strength and $\tilde{V}(\pp) = F_\perp \left(  \pp \right) \cos^2 \alpha + F_\parallel (\pp) \sin^2 \alpha$ is the quasi-2D momentum-space dipolar interaction potential, with $F_\perp(\pp) = 2 - 3 \sqrt{\pi} p e^{p^2} \mathrm{erfc}[p]$ and $F_\parallel(\pp) = -1 + 3 \sqrt{\pi} (p_y^2 / p) e^{p^2} \mathrm{erfc}[p]$.  Here, $\mathrm{erfc}[p]$ is the complimentary error function and $\alpha$ is the polarization tilt angle between $\dd = d ( \hat{z} \cos \alpha  +  \hat{y} \sin \alpha) $ and the $z$-axis.  The Beliaev damping rate for a quasiparticle with momentum $\pp$ is found to be
\begin{align}
\label{Beliaev}
\gamma_{\mathrm{B},\pp} = \frac{2\pi}{\hbar}  \sum_{\kk \qq} | \bar{ \mathcal{A}}^\pp_{\kk  \qq} |^2 \delta ( \omega_\pp - ( \omega_\kk + \omega_{\qq} ) ),
\end{align}
where  $ \bar{\mathcal{A}}^\pp_{\kk  \qq} = \mathcal{A}^\pp_{\kk \qq} + \mathcal{A}^\pp_{\qq \kk}$, and $\mathcal{A}^\pp_{\kk \qq}$ has matrix elements
\begin{align}
&\mathcal{A}^\pp_{\kk \qq} = \pi \sqrt{n_0} \left[ u_\pp \left( \tilde{V}(\kk) (u^*_\kk u^*_\qq + v^*_\kk u^*_\qq) + \tilde{V}(\kk+\qq) v^*_\kk u^*_\qq   \right)  \right. \nonumber \\
&+ \left. v_\pp \left( \tilde{V}(\kk) (v^*_\kk v^*_\qq + u^*_\kk v^*_\qq) + \tilde{V}(\kk+\qq) u^*_\kk v^*_\qq   \right)   \right] \delta_{\pp, \kk+\qq}.
\end{align}
We take the thermodynamic limit, and evaluate Eq.~(\ref{Beliaev}) numerically. 

\begin{figure}[t]
\includegraphics[width=.9\columnwidth]{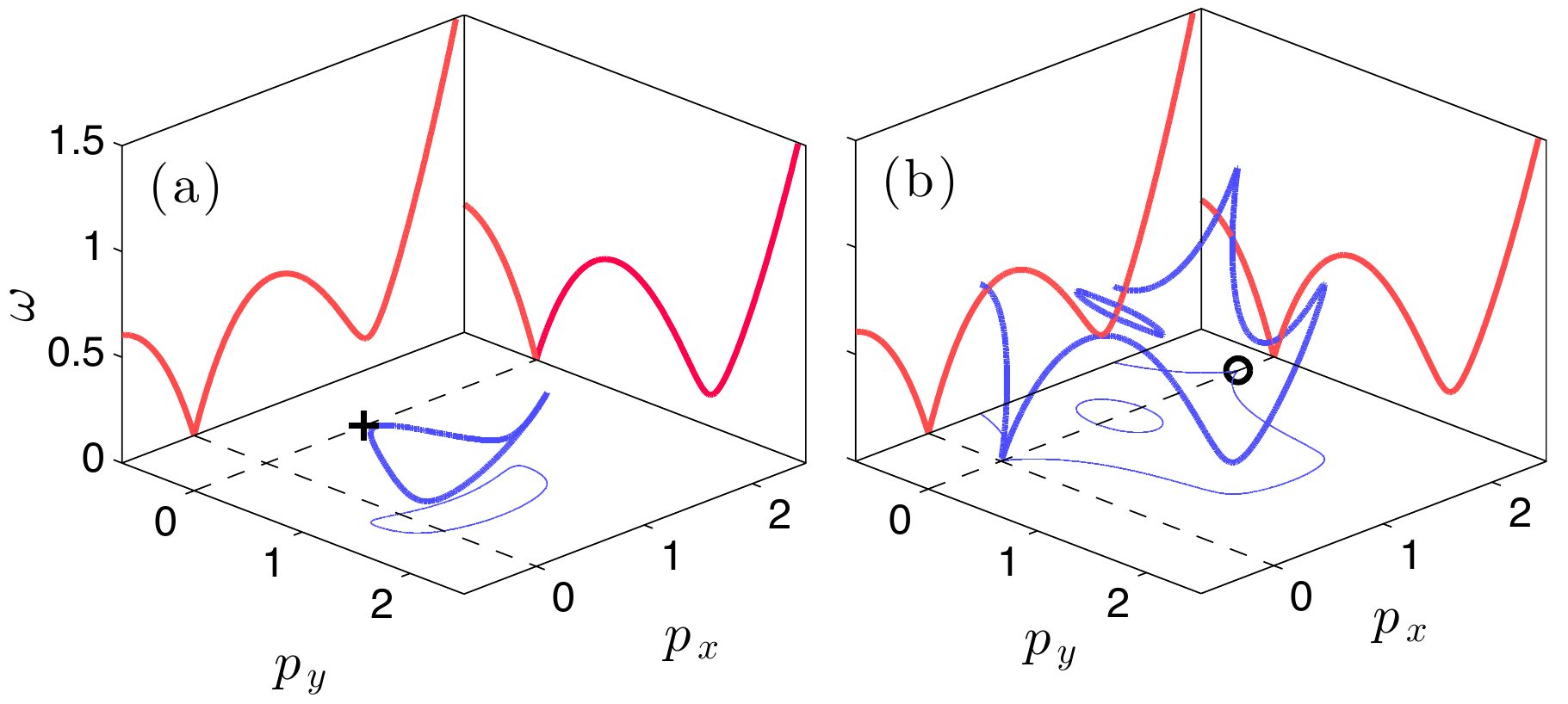}
\caption{\label{fig:en} (color online).  Energetics of Beliaev damping for a quasiparticle with momentum (a) $\pp l_z = 0.9 \hat{x} $, a maxon (black $+$), and (b) $\pp l_z = 2.2 \hat{x} $, in the free-particle part of the spectrum (black circle), in a quasi-2D dipolar condensate with $n_0 g_d = 1.7 $ and $\alpha=0$.  The vertical panels show the quasiparticle spectrum (red lines).  The dark blue lines show the manifold of decay channels allowed by energy and momentum conservation.  The light blue lines show the allowed momenta only.   }
\end{figure}

We first consider a quasi-2D dipolar condensate that is polarized perpendicular to the 2D plane ($\alpha=0$).  In this case, an expansion of the small-momentum, phonon part of the quasiparticle spectrum gives $\omega_\pp \simeq c_d p (1 - \sqrt{ 9\pi / 32} p + \ldots)$, where $c_d = \sqrt{2 n_0 g_d}$ is the phonon speed. This downward curvature prohibits the Beliaev damping of phonons, due to the impossibility of simultaneous energy and momentum conservation.  Thus, phonons do not damp in quasi-2D dipolar condensates~\cite{Natu13a}. This is in contrast to quasi-2D condensates with repulsive, isotropic contact interactions, which host quasiparticle spectra with upward curvature at small momenta, resulting in Beliaev damping rates $\propto p^3$ at small $p$~\cite{Chung09}.

\begin{figure}[t]
\includegraphics[width=.9\columnwidth]{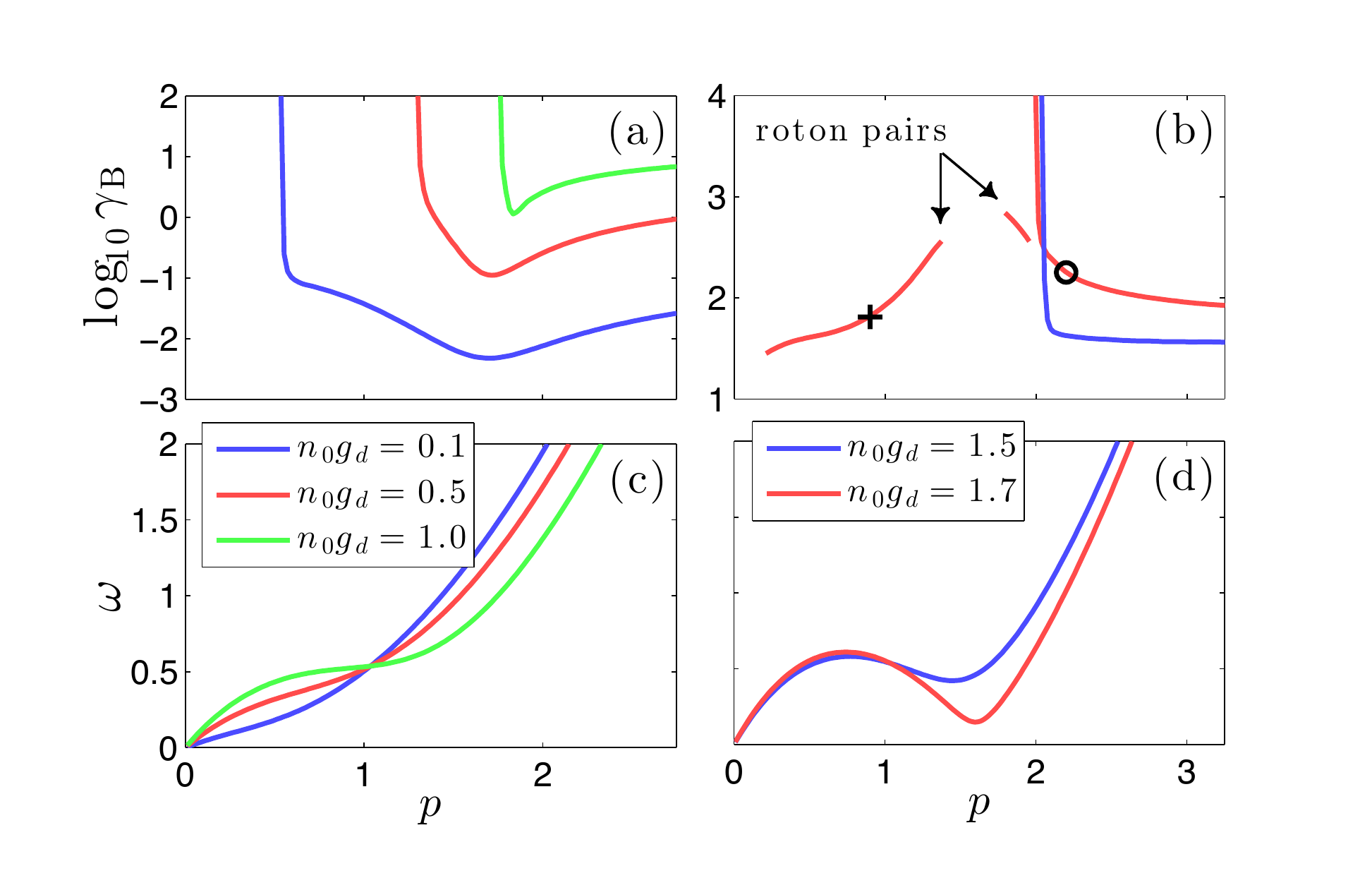}
\caption{\label{fig:rates} (color online). (a) Beliaev damping rate for weak dipolar interactions, where no roton is present in the quasiparticle spectrum; the corresponding spectra are shown in (c).  (b) Rate for stronger dipolar interactions, where a roton is present; the corresponding spectra are shown in (d).   }
\end{figure}

At larger momenta, a local ``roton'' minimum with an energy gap $\Delta_\mathrm{r}$ develops in the quasiparticle spectrum for dipolar interaction strengths $n_0 g_d \gtrsim 1.15$, which ultimately softens to $\Delta_\mathrm{r} = 0$ at a momentum $ p_\mathrm{r} \simeq 1.62$ when $n_0 g_d \simeq 1.72$.  This is accompanied by a local ``maxon'' maximum at $p  \simeq 0.74$.  An example roton-maxon spectrum for $n_0 g_d = 1.7$ is shown in the vertical panels of Fig.~\ref{fig:en} and by the red curve in Fig.~\ref{fig:rates}(d).

As the roton minimum develops, the density of quasiparticle states grows significantly.  Near the minimum, the spectrum can be expanded about $p \sim p_\mathrm{r}$ to give $\omega_\pp \simeq \Delta_\mathrm{r} +  (p-p_\mathrm{r})^2 / 2 m_\mathrm{r} $ where $m_\mathrm{r}$ is the effective roton mass.  The density of states near the roton minimum is thus $\rho_\mathrm{r} (\omega ) = 2 \pi m_\mathrm{r}  (1 + p_\mathrm{r} / \sqrt{2 m_\mathrm{r} (\omega - \omega_\mathrm{r}) } )$.  The divergence of this expression at $\omega = \omega_\mathrm{r}$ contributes to an anomalously large density of states in this vicinity.  It is instructive to note that the expression for the Beliaev damping rate in Eq.~(\ref{Beliaev}) is reminiscent of Fermi's Golden Rule, which describes the scattering of a quantum state into other final states at a rate proportional to the density of available final states.  Indeed, the evaluation of the Dirac-delta function in Eq.~(\ref{Beliaev}) produces a factor resembling the density of final quasiparticle states; we thus expect large damping rates for quasiparticles that can decay into rotons.

In Fig.~\ref{fig:en}, we illustrate the manifold of available final quasiparticle states for $n_0 g_d = 1.7$, which supports a prominent roton-maxon feature.  In panel (a), we consider a quasiparticle with momentum $\pp = 0.9 \hat{x} $ (shown by the black + sign in the $p_x$-$p_y$ plane), which is in the maxon part of the spectrum.  Whenever the maxon energy exceeds $2 \Delta_\mathrm{r}$, it is energetically possible to decay into a pair of rotons.  The blue lines, which show the energy and momenta of the available final quasiparticle states, are centered about the roton minima in the $+y$ and $-y$ directions.  This indicates that maxons undergo Beliaev damping by decaying into a pair of nearly counter-propagating rotons that travel transverse to the initial quasiparticle direction.  In panel (b) of Fig.~\ref{fig:en}, we consider a quasiparticle with momentum $\pp = 2.2 \hat{x} $ (black circle), which is in the higher energy, free particle-like part of the spectrum.  A number of final states are available to these quasiparticles; they can decay into rotons, maxons, and phonons. In the latter process, the quasiparticle ``sheds''  low-energy phonons and loses a correspondingly small amount of energy and momentum.  In the former processes, many combinations are final states are possible. Roton-maxon pairs can be produced, or pairs of forward-propagating quasiparticles; the momenta of these final states are shown by the detached blue loop in Fig.~\ref{fig:en}(b).  

We plot the Beliaev damping rates for a range of dipolar interaction strengths in Fig.~\ref{fig:rates}; the rates are scaled by the axial trap frequency $\omega_z$.  Panel (a) shows rates for quasiparticle spectra that lack roton-maxon features (shown in panel (c)).   The downward curvature of the quasiparticle spectrum forbids phonon damping below a critical momentum $p_\mathrm{crit}$; this is apparent in all cases shown.  For small $p$, $p_\mathrm{crit} = \sqrt{9 \pi} c_d^2$,  and $p_\mathrm{crit} \sim c_d$ for larger $p$ \cite{Natu13a}. Additionally, we see evidence that as $n_0 g_d$ increases, the downward curvature of the quasiparticle spectrum becomes more pronounced, and $p_\mathrm{crit}$ increases correspondingly.  As $p \rightarrow p_\mathrm{crit}$ from above, the damping rate becomes anomalously large.   The only available mechanism for Beliaev damping in this small range of momenta near $p_\mathrm{crit}$ is the shedding of low energy phonons.  We attribute these anomalously large damping rates to the unique curvature of the quasiparticle spectrum, which produces a factor resembling a large density of phonon states in the evaluation of Eq~(\ref{Beliaev}).

Note that for $n_0 g_d = 0.1$, the Beliaev damping rates are very small, remaining much less than $\omega_z$ in the range of $p$ shown.  These rates are nearly an order of magnitude smaller than those of a quasi-2D \emph{non}-dipolar BEC with an equivalent chemical potential.  As $n_0 g_d$ increases, the damping rates increase significantly across the range of $p$, which we expect due to the proportionality $\gamma_\mathrm{B} \propto g_d^2$.  The rates at larger $p$ become more comparable to those of a system with purely contact interactions as $n_0 g_d \rightarrow 1$.  

Panel (b) of Fig.~\ref{fig:rates} shows Beliaev damping rates for larger dipolar interaction strengths, which support spectra with pronounced roton-maxon features (shown in panel (d)).  We consider two distinct cases; for $n_0 g_d = 1.5$, the maxon energy is less than $2 \Delta_\mathrm{r}$ (blue curve), and for $n_0 g_d = 1.7$, the maxon energy is greater than $2 \Delta_\mathrm{r}$ (red curve).  In the former case, it is energetically forbidden for a maxon to decay into a pair of rotons.  Thus, \emph{all} low-energy quasiparticles (phonons, maxons, and rotons) remain undamped, and Beliaev damping only occurs for $p \gtrsim 2$.  For the latter case, a maxon can damp into a pair of transverse, counter-propagating rotons.  Notice that the red curve in panel (b) is separated into three distinct parts.  The two left-most parts correspond to Beliaev damping into roton pairs only.  These damping rates are anomalously large, achieving values well over $100 \omega_z$ for some values of $p$.  This is due to the large density of states near the roton minimum.  Additionally, the Beliaev damping rate vanishes for a range of $p$ near the roton minimum, reflecting the fact that rotons are \emph{undamped}, due to their anomalously low energy and large momenta.
The black $+$ sign and circle show the Beliaev damping rates for quasiparticles with $p=0.9$ and $p=2.2$ respectively, corresponding to the discussion of Fig.~\ref{fig:en}.
Though the non-condensate density grows as $\Delta_\mathrm{r}$ softens~\cite{Fischer06}, it remains dilute for the cases we consider here.  We thus expect our perturbation theory to remain valid for these large damping rates.

By tilting the external polarizing field off axis ($\alpha \neq 0$), the dipolar interactions can be made strongly anisotropic.  It is predicted that anisotropic dipolar interactions will produce a quasiparticle spectrum with correspondingly strong anisotropies, supporting rotons for only a narrow range of propagation directions~\cite{Ticknor11}.  Such anisotropic spectra have important consequences for the damping of quasiparticles in these systems.  

\begin{figure}[t]
\includegraphics[width=.9\columnwidth]{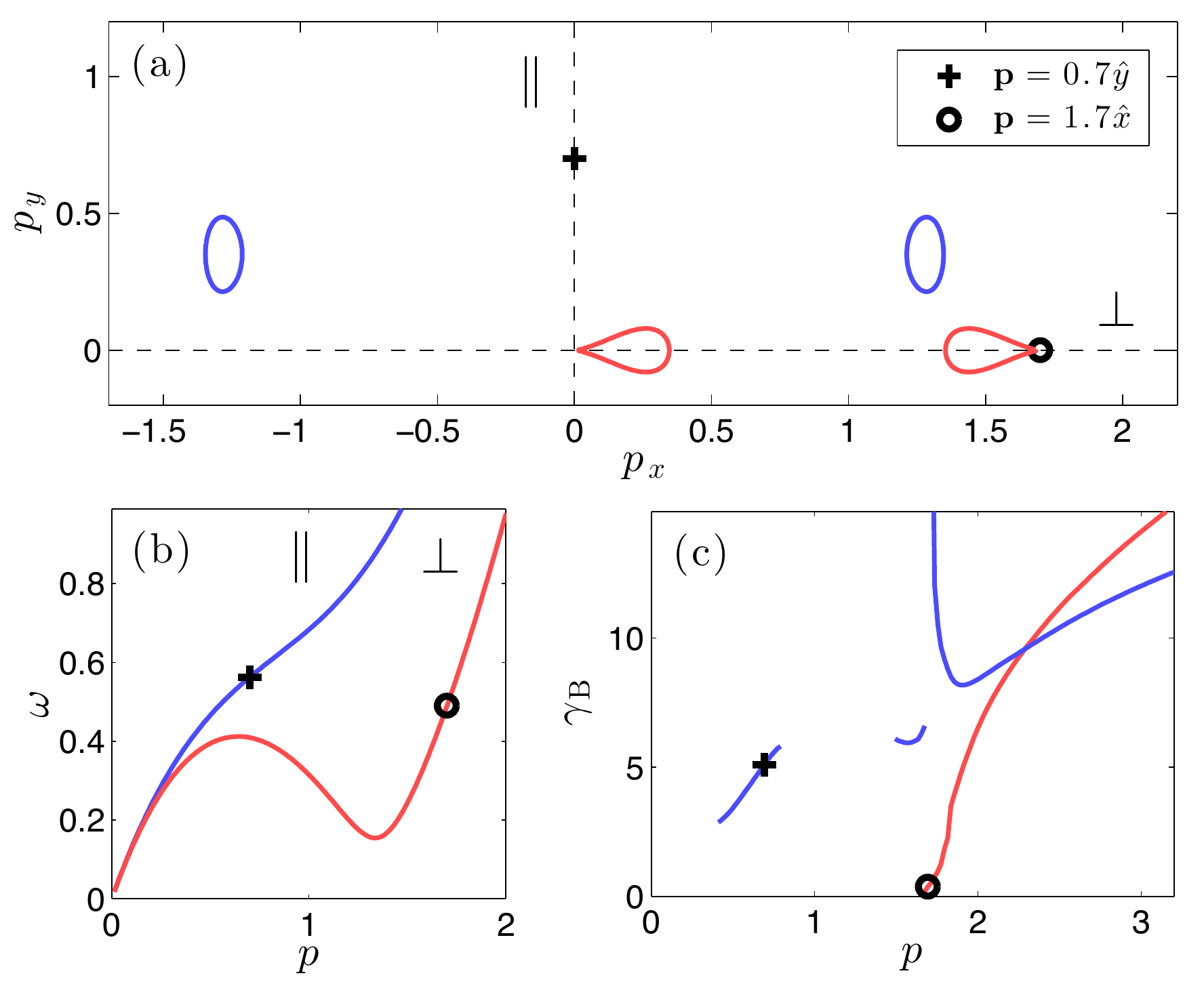} \\
\caption{\label{fig:aniso} (color online).  Beliaev damping of a quasi-2D dipolar condensate with $n_0 g_d = 1.3$, $\alpha = \pi / 8$, and $\theta = 0$ (dipoles tilted in the $y$-direction). (a) Manifold of allowed decay channels for a quasiparticle with momentum $\pp  = 0.7 \hat{y} $ (black +) shown by the blue lines, and momentum $\pp = 1.7 \hat{x} $ (black circle) shown by the red lines.  (b) Spectrum for quasiparticles propagating in the $y$-direction (blue line) and the $x$-direction (red line).  (c)  The corresponding Beliaev damping rates.  }
\end{figure}

We plot the quasiparticle spectra and Beliaev damping rates for a condensate with $n_0 g_d = 1.3$ and a tilt angle $\alpha= \pi/8$ in Figs.~\ref{fig:aniso}(b) and~\ref{fig:aniso}(c), respectively.  In this case, the spectrum for quasiparticles propagating in the $x$-direction ($\perp$, red line) exhibits roton-maxon character, while the spectrum in the $y$-direction ($\parallel$, blue line) does not.  Above, we noted that maxons can only undergo Beliaev damping by decaying into a pair of nearly counter propagating rotons when $\alpha=0$.  Here, no rotons exist in the transverse direction, and maxons are consequently \emph{undamped} despite the fact that the maxon energy exceeds $2\Delta_\mathrm{r}$.  Quasiparticles propagating in the $x$-direction begin to damp near $\pp = 1.7 \hat{x}$, shown by the black circle(s) in Fig.~\ref{fig:aniso}.  The momenta of the available final states are shown by the red line in panel (a), and the corresponding damping rates are shown in panel (c).  The onset of damping is due to the shedding of phonons near this momentum. Interestingly, the damping rate is not anomalously large near this onset, unlike the $\alpha = 0$ case; this is due to the anisotropy of the spectrum, which skews its curvature unfavorably.

Although no roton-maxon feature exists in the $y$-direction, quasiparticles propagating in this direction can damp by  decaying into \emph{transverse} roton pairs, as illustrated by the blue lines in Fig.~\ref{fig:aniso}(a), which show the momenta  of the available final states for a quasiparticle with $\pp = 1.7 \hat{x} $ (shown by the black + sign). The damping rates for this process are shown by the two left-most blue line segments in panel (c). For larger momenta, the quasiparticles begin to shed phonons in the $y$-direction.  The damping rates for these momenta are shown by the right-most blue line in panel (c).  Thus,  
the anisotropic dipolar interactions not only lead to anisotropic damping rates, but rather qualitatively different damping mechanisms depending on the direction of quasiparticle propagation.  

Our predictions have important consequences for ongoing experiments with ultracold dipolar atoms.  For example, in experiments measuring the dynamic structure factor $S(\pp,\omega)$ of the condensate via, for example, optical Bragg scattering~\cite{Kozuma99}, these rates will appear as spectral widths~\cite{Blakie12}. Further, our results can be used to predict the short-time non-equilibrium dynamics of these systems, as the Beliaev mechanism is responsible for the redistribution of quasiparticles near $T=0$.  Take, for example, the anisotropic case discussed above.  If an oblate dipolar condensate is prepared with $n_0 g_d = 1.3$ and $\alpha = \pi/8$, and modes with $\pp = 0.5 \hat{x}$ are excited, they should undergo coherent dynamics for long times.  On the other hand, the excitation of modes with $\pp = 0.5 \hat{y}$ will result in the nearly immediate redistribution of energy into transverse rotons.  In this sense, the anisotropic Beliaev damping should result in strongly anisotropic relaxation dynamics.  Experimentally, the limit of a deep roton ($n_0 g_d = 1.7$) can be achieved, for example, with $^{164}$Dy~\cite{Lu11} in an oblate trap with axial frequency $\omega_z = 2\pi \times 10^3 \, \mathrm{Hz}$ and a mean 3D density $\bar{n}_\mathrm{3D} \sim 3 \times 10^{14} \, \mathrm{cm^{-3}}$.  For strongly dipolar molecules, much smaller densities are required. 

\acknowledgements{\emph{Acknowledgements ---} We thank John Corson for his invaluable correspondence.  RW acknowledges partial support from the Office of Naval Research under Grant No.~N00014115WX01372, and from the National Science Foundation under Grant No.~PHY-1516421.  SN acknowledges support from LPS-CMTC, LPS-MPO-CMTC, AROs Atomtronics MURI, and the AFOSRs Quantum Matter MURI.

\end{document}